\documentclass[aps, prb, twocolumn, superscriptaddress, letterpaper]{revtex4-2}

\usepackage{graphicx}
\usepackage{amssymb}
\usepackage{amsmath}
\usepackage{txfonts}
\usepackage{bm}
\usepackage{color}
\usepackage[colorlinks=true, linkcolor=blue, anchorcolor=black, citecolor=blue, filecolor=black, menucolor=black, runcolor=black, urlcolor=blue]{hyperref}
\usepackage{orcidlink}

\newcommand{\SI}{Supplementary Information}

\begin{document}

\title{Observation of returning Thouless pumping}

\author{Zheyu Cheng\orcidlink{0000-0001-5009-7929}}
\affiliation{Division of Physics and Applied Physics, School of Physical and Mathematical Sciences, Nanyang Technological University,
Singapore 637371, Singapore}

\author{Sijie Yue}
\affiliation{National Laboratory of Solid State Microstructures and Department of Physics, Nanjing University, Nanjing 210093, China}

\author{Yang Long\orcidlink{0000-0001-7600-3396}}
\affiliation{Division of Physics and Applied Physics, School of Physical and Mathematical Sciences, Nanyang Technological University,
Singapore 637371, Singapore}

\author{Wentao Xie}
\affiliation{Department of Physics, The Chinese University of Hong Kong, Shatin, Hong Kong SAR, China}

\author{Zixuan Yu}
\affiliation{Division of Physics and Applied Physics, School of Physical and Mathematical Sciences, Nanyang Technological University,
Singapore 637371, Singapore}

\author{Hau Tian Teo\orcidlink{0000-0003-4488-5915}}
\affiliation{Division of Physics and Applied Physics, School of Physical and Mathematical Sciences, Nanyang Technological University,
Singapore 637371, Singapore}

\author{Y. X. Zhao\orcidlink{0000-0002-3084-1033}}
\email{yuxinphy@hku.hk}
\affiliation{Department of Physics and HK Institute of Quantum Science \& Technology, The University of Hong Kong, Pokfulam Road, Hong Kong, China}

\author{Haoran Xue\orcidlink{0000-0002-1040-1137}}
\email{haoranxue@cuhk.edu.hk}
\affiliation{Department of Physics, The Chinese University of Hong Kong, Shatin, Hong Kong SAR, China}
\affiliation{State Key Laboratory of Quantum Information Technologies and Materials, The Chinese University of Hong Kong, Shatin, Hong Kong SAR, China}

\author{Baile Zhang\orcidlink{0000-0003-1673-5901}}
\email{blzhang@ntu.edu.sg}
\affiliation{Division of Physics and Applied Physics, School of Physical and Mathematical Sciences, Nanyang Technological University,
Singapore 637371, Singapore}
\affiliation{Centre for Disruptive Photonic Technologies, Nanyang Technological University, Singapore 637371, Singapore}

\begin{abstract}
Introduced by David Thouless in 1983, Thouless pumping exemplifies topological properties in topological systems, where the transported charge is quantized by the Chern number. Recently, returning Thouless pumping was theoretically proposed, in which quantized charge is pumped during the first half of the cycle but returns to zero in the second half. This mechanism leads to crystalline symmetry-protected delicate topological insulators. Unlike conventional topological bands, a delicate topological band is Wannierizable but not atomically obstructed, which features multicellular Wannier functions extending beyond a single unit cell. Here, by replacing the second dimension with a synthetic dimension, we realize a two-dimensional delicate topological insulator via a set of one-dimensional acoustic crystals with fine-tuned geometric parameters. Through acoustic bands and wavefunction measurements, we directly observe returning Thouless pumping and symmetric multicellular Wannier functions, followed by establishing the bulk-boundary correspondence between sub-Brillouin zone Chern numbers and gapless boundary modes. As enriched by crystalline symmetries, our experimental demonstration of returning Thouless pumping expands the current understanding of topological phases of matter.
\end{abstract}

\maketitle

\textit{Introduction.}---Thouless pumping describes the quantized transport of charge through an adiabatic and cyclic evolution of a Hamiltonian, occurring in the absence of a net external bias~\cite{thouless1983quantization}. The transported charge is quantized by the system's Chern number, providing topological protection against perturbations that are small compared to the energy gap between the ground and excited states~\cite{xiao2010berry, citro2023thouless}. As a fundamental topological phenomenon, Thouless pumping has been extensively studied both theoretically~\cite{niu1990towards, fu2022nonlinear, wang2013topological, ke2016topological, jurgensen2022chern, fu2022two, ravets2025thouless, long2019floquet} and experimentally~\cite{ma2018experimental, lohse2016thouless, nakajima2016topological, viebahn2024interaction, jurgensen2021quantized, fedorova2020observation, cerjan2020thouless, cheng2020experimental, you2022observation, grinberg2020robust} across diverse physical platforms, including electronic systems~\cite{niu1990towards, ma2018experimental, fu2022nonlinear}, ultracold atoms~\cite{wang2013topological, lohse2016thouless, nakajima2016topological, viebahn2024interaction}, photonics~\cite{ke2016topological, jurgensen2022chern, jurgensen2021quantized, fedorova2020observation, cerjan2020thouless, fu2022two, ravets2025thouless}, acoustics~\cite{long2019floquet, cheng2020experimental, you2022observation}, and mechanics~\cite{grinberg2020robust}.

Previously focused on global symmetries~\cite{citro2023thouless}, recent studies have shown that crystalline symmetry can generalize Thouless pumping~\cite{alexandradinata2021teleportation, nelson2021multicellularity, nelson2022delicate}. In these scenarios, the Chern number over the entire Brillouin zone (BZ) remains zero, meaning that no net charge is pumped over an adiabatic cyclic evolution. However, when the BZ is partitioned into multiple sub-Brillouin zones (sBZs) via its high-symmetry lines, each sBZ can carry a nonzero Chern number. As the adiabatic parameter evolves through a full cycle, the bulk polarization evolves from 0 to 1 then returns to 0, effectively realizing two consecutive but opposite Thouless pumps. The large variation in bulk polarization is characterized by a $2\pi$ quantum of Berry-Zak phase difference between two high-symmetry lines, thereby defining the returning Thouless pump (RTP).

RTP can significantly change the nature of Wannier functions (WFs)---the Fourier transforms of Bloch functions---revealing that they necessarily extend beyond a single unit cell. This topological obstruction prevents WFs from being adiabatically deformed into a single primitive unit cell while preserving all crystalline symmetries~\cite{alexandradinata2021teleportation, nelson2022delicate, chen2023stably}. However, despite this obstruction, exponentially localized WFs can still be constructed in a way that respects crystalline symmetries~\cite{alexandradinata2018no}. This phenomenon, known as multicellular topology, gives rise to a new type of topological insulator---delicate topological insulator~\cite{nelson2021multicellularity}.

Unlike conventional topological bands, delicate topological insulators~\cite{nelson2021multicellularity} fall beyond existing classifications, signifying that topological materials are not limited to obstructed insulators~\cite{haldane1988model, kane2005quantum, bernevig2006quantum} and fragile topology~\cite{po2018fragile}. As the name suggests, delicate topology is even more vulnerable than fragile topology; it can be trivialized simply by adding additional bands to either the valence or conduction bands. These distinctive properties have raised intense theoretical discussions~\cite{nelson2021multicellularity, schindler2021noncompact, lapierre2021n, nelson2022delicate, chen2023stably, zhu2023scattering, graf2023massless, zhu2023z, lim2023real, brouwer2023homotopic, chen2024chern}, yet an experimental realization of delicate topology remains an open question.

\begin{figure*}
  \centering
  \includegraphics[width=\textwidth]{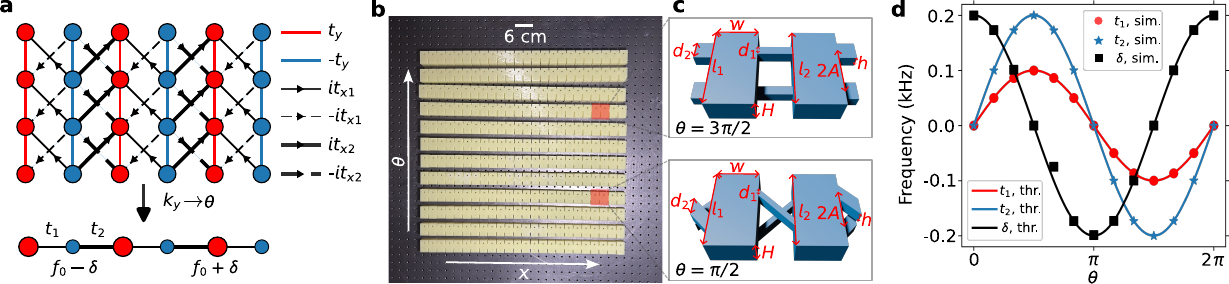}
\caption{Design of a synthetic acoustic delicate topological insulator.  \textbf{a}, Upper panel: A 2D tight-binding model of a delicate topological insulator, where the red and blue circles denote the two sublattices. Lower panel: By mapping the momentum $k_y$ to a parameter $\theta$, the 2D model can be reduced to a set of 1D models parameterized by $\theta$.  \textbf{b}, Photo of the fabricated acoustic crystal samples with $\theta$ gradually changing from 0 to $2\pi$. \textbf{c}, Unit cells of the acoustic crystals with $\theta = 3\pi/2$ (upper panel) and $\theta = \pi/2$ (lower panel). \textbf{d}, Plots showing the relationship between the tight-binding parameters and $\theta$. The curves and scatters correspond to the ideal values in the tight-binding model and the values extracted from the acoustic model, respectively. }
  \label{fig1}
\end{figure*}

The main challenge in realizing delicate topological phases stems from theoretical models that typically require long-range and complex-valued hoppings~\cite{alexandradinata2021teleportation, nelson2021multicellularity, nelson2022delicate}. Here, by mapping a momentum component to the synthetic dimension~\cite{yuan2018synthetic, ozawa2019topological2}, we design a $1+1$D acoustic system with only nearest-neighbor and real-valued hoppings, which corresponds to a two-dimensional delicate topological insulator with long-range and complex-valued hoppings. We observe the RTP by computing bulk polarization using experimentally measured wavefunctions. Furthermore, we observe a symmetric multicellular WF after selecting a gauge obeying the crystalline symmetry. We also establish the bulk-boundary correspondence of delicate topology through the observation of gapless edge modes protected by sBZ Chern numbers.

\textit{Design of a synthetic acoustic structure.}---To realize the RTP, we first consider the following tight-binding model for a 2D delicate topological insulator~\cite{zhu2023scattering}:
\begin{align}
H (k_x, k_y) =& 2\left( {{t_{x1}} + {t_{x2}}\cos {k_x}} \right){\sin {k_y}}{\sigma _x} + 2{t_{x2}}{\sin {k_y}\sin {k_x}} {\sigma _y} \notag\\
&+ 2{t_y}\cos {k_y}{\sigma _z}	,\label{eq01}
\end{align}
where $t_{x1}$, $t_{x2}$ and $t_y$ are real hopping parameters, $\sigma_{x,y,z}$ are Pauli matrices. This model respects a mirror symmetry $\mathcal M_y = \sigma_z$ along the $y$ direction, and a time reversal symmetry $\mathcal T = \sigma_z \mathcal K$ where $\mathcal K$ is the complex conjugation operator (see \SI).  

The high complexity of this model can be readily appreciated from its real-space representation, shown in Fig.~\ref{fig1}a, which includes both next-nearest neighbor imaginary hoppings (e.g., $it_{x1}$) and negative hoppings (e.g., $-t_y$). While negative hoppings can be engineered in wave systems with careful designs~\cite{keil2016universal, serra2018observation, peterson2018quantized, xue2020Observation, ni2020demonstration, xue2022projectively, li2022acoustic}, implementing imaginary hoppings is highly challenging. To address this issue, we adopt a dimension reduction procedure by mapping the momentum $k_y$ to a synthetic parameter $\theta$. Consequently, at each fixed $\theta$, the 2D lattice model reduces to a 1D one without any imaginary hoppings (see the lower panel of Fig.~\ref{fig1}a), whose Hamiltonian takes the form:
\begin{equation}
H = \left( {{t_1} + {t_2}\cos {k_x}} \right){\sigma _x} + {t_2}\sin {k_x}{\sigma _y} + \delta {\sigma _z}	,\label{eq02}
\end{equation}
with
\begin{equation}
{t_1} = 2t_{x1}\sin \theta ,\;{t_2} = 2t_{x2}\sin \theta ,\;\delta  = 2t_y\cos \theta.
\label{eq03}
\end{equation}
Here, we treat $\theta$ purely as a synthetic momentum, and the physics of the 2D model is accessed by studying a series of 1D models with different values of $\theta$.

We design an acoustic crystal to realize the $1+1$D lattice model (i.e., Eq.~\eqref{eq02}). A photograph of the samples with different values of $\theta$ is shown in Fig.~\ref{fig1}b, where the unit cell (see Fig.~\ref{fig1}c) consists of two acoustic resonators coupled via thin channels with rectangular cross-sections. The samples are hollow and bounded by rigid walls, with two holes on each resonator for excitation and detection. Each resonator supports a dipole mode within the frequency range of interest (i.e., 5.3--6.1 kHz), which plays the role of an atomic orbital in the tight-binding model. Due to the dipolar nature of the mode, we can control the hopping sign by connecting different parts across the mode's nodal line, then adjust the hopping amplitude by tuning the position and width of the hopping channel~\cite{xue2020Observation, ni2020demonstration, xue2022projectively, li2022acoustic}.

To design a set of 1D acoustic crystals parameterized by $\theta$, we numerically compute the acoustic dispersions for different geometric parameters (i.e., heights of the resonators, widths and tilting angles of the hopping channels) and subsequently fit the results to the tight-binding model. Through an exhaustive parameter scan, we obtain a dataset containing a detailed correspondence between the structural parameters and tight-binding parameters. Using this dataset, we can easily identify a series of structures that continuously change with $\theta$ and satisfy the conditions given in Eq.~\eqref{eq03}, as shown in Fig.~\ref{fig1}d (see \SI for more structural details). Notably, for positive hoppings (i.e., $0 < \theta < \pi$), the hopping channels are crossed (see the lower panel of Fig.~\ref{fig1}c), while for negative hoppings (i.e., $\pi < \theta < 2\pi$), the hopping channels are parallel to each other (see the upper panel of Fig.~\ref{fig1}c). In the experiment,  we adopt $t_{x1} = 0.05$ kHz, $t_{x2} = t_y = 0.1$ kHz, and $f_0 = 5.7$ kHz, and we use 12 discrete values of $\theta$ in our subsequent experiments.

\begin{figure*}
  \centering
  \includegraphics[width=\textwidth]{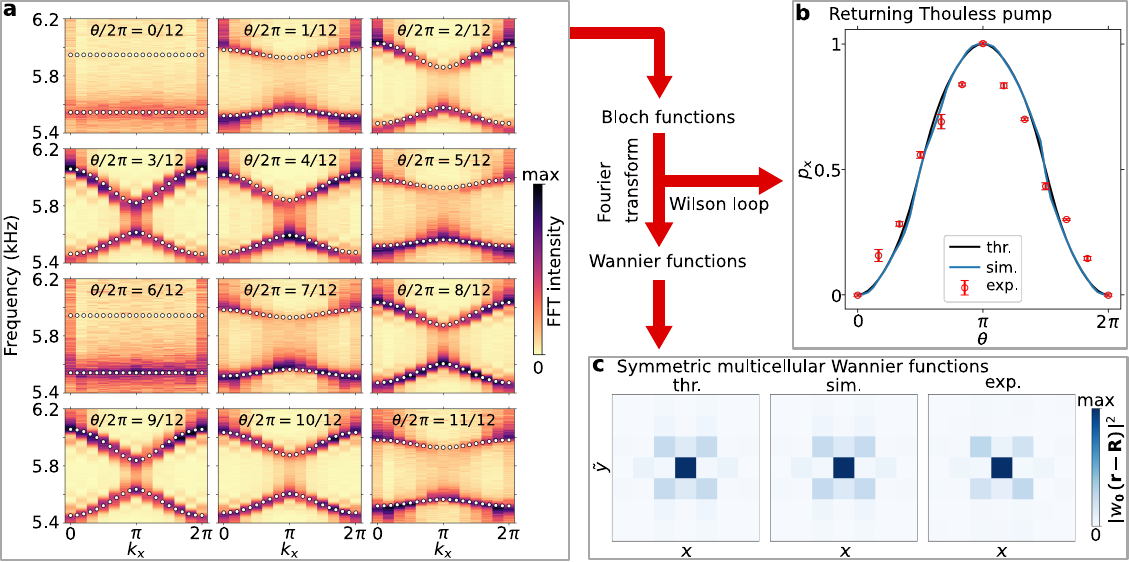}
\caption{Observation of the returning Thouless pump and symmetric multicellular Wannier functions.  \textbf{a}, Measured (colormaps) and simulated (white dots) bulk dispersions for different values of $\theta$. \textbf{b}, Plots of bulk polarization against $\theta$. The black and blue curves correspond to the results obtained from tight-binding calculations and full-wave simulations, respectively. The experimental results are represented by the red dots, with the error bars denoting the standard deviation from five independent measurements.  \textbf{c}, Plots of symmetric multicellular Wannier functions of the lower band. Here, $\tilde{y}$ is the synthetic spatial dimension corresponding to the synthetic momentum $\theta$.}
  \label{fig2}
\end{figure*}

\textit{Observation of the RTP.}---RTP exhibits two Thouless pumping processes in opposite directions as momentum traverses the BZ, with the pumped charge quantized in the first half-period. Here, we provide direct validation of this unique phenomenon through extensive wavefunction measurements. Excited by an acoustic point source, we measure the field distributions and Fourier-transform the real-space fields to obtain the wavefunctions in momentum space (see \SI). In our $1+1$D acoustic crystal, an advantage is that we can fix the synthetic momentum $\theta$ and perform individual experiments in 1D systems, allowing us to resolve the other physical momentum $k_x$ more efficiently, with modes at different parts of the BZ being easier to be excited (see Methods). Consequently, the full information about band dispersion and wavefunctions can be obtained by repeating the measurements on multiple samples with different values of $\theta$. Under this scheme, the resolutions of the physical and synthetic momenta are restricted by the number of unit cells in each 1D sample and the number of 1D samples, respectively. Practically, we find that 12 samples with 12 unit cells each are sufficient to arrive at accurate results.

Figure~\ref{fig2}a shows the measured bulk dispersions of the 12 samples, which match well with the numerical results obtained from full-wave simulations (indicated by the white dots). In the cases of $\theta=0$ and $\theta=\pi$, the two resonators become decoupled (see Fig.~\ref{fig1}d), and we choose to excite only the lower bands by placing the source in the resonator with a lower resonant frequency. As a result, the upper band is not visible in these two cases. For the other samples, we also place the source in the resonator with a lower resonant frequency to ensure efficient excitation of the modes in the lower band. The wavefunctions can then be constructed by extracting the sublattice-resolved Fourier fields at the measured frequencies of the lower band. The bulk polarization at each fixed $\theta$ is evaluated by calculating the Berry phase through a discrete Wilson loop~\cite{zak1989berry}:
\begin{equation}
{p_x}\left( \theta  \right) =  - \frac{i}{{2\pi }}\ln \left[ {\left\langle {{u_{{k_N}}}|{u_{{k_{N - 1}}}}} \right\rangle \left\langle {{u_{{k_{N - 1}}}}|{u_{{k_{N - 2}}}}} \right\rangle  \cdots \left\langle {{u_{{k_2}}}|{u_{{k_1}}}} \right\rangle } \right], \label{Berry_phase}
\end{equation}
where $|u_{k_i}\rangle$ is the measured wavefunction (of the lower band) at momentum $k_i=2\pi{i}/{N}$, with $N=12$ (we set $|u_{k_{12}}\rangle\equiv|u_{k_0}\rangle$ to ensure a periodic gauge). As shown in Fig.~\ref{fig2}b, the measured bulk polarization continuously changes from 0 to 1 as $\theta$ traverses half of the BZ and returns to 0 during the other half of the evolution. To the best of our knowledge, this is the first experimental observation of the RTP. For comparison, we also plot the results from full-wave simulations and the tight-binding model in Fig.~\ref{fig2}b, both of which agree well with the experiment. 

\begin{figure*}
  \centering
  \includegraphics[width=0.9\textwidth]{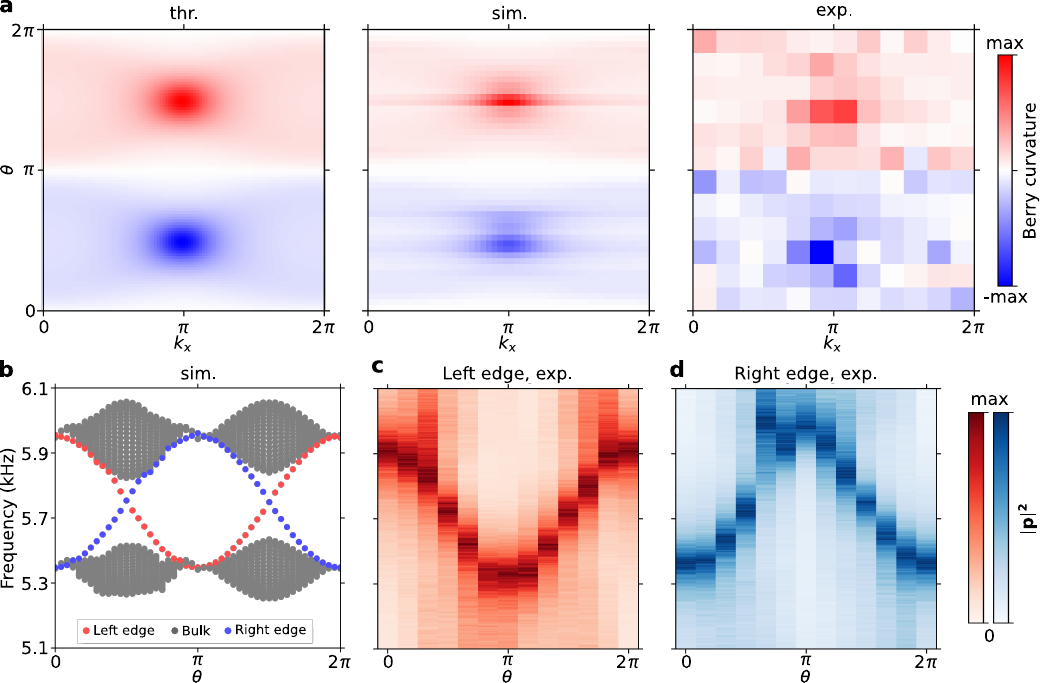}
\caption{Observation of multiple gapless edge modes. \textbf{a}, Distribution of the Berry curvature in the 2D Brillouin zone in the tight-binding model, full-wave simulation, and experiment, respectively. The lower and upper halves of the Brillouin zone have opposite Berry curvatures, leading to sub-Brillouin zone Chern numbers of $-1$ and $+1$, respectively.  \textbf{b}, Plot of eigenfrequencies for finite-size acoustic crystals with different $\theta$. The grey dots denote bulk modes, while the red and blue dots correspond to edge modes at the left and right edges, respectively.  \textbf{c(d)}, Measured acoustic intensities at the left (right) edges. The source and probe are placed in the same outermost resonator in the measurement.}
  \label{fig3}
\end{figure*}

\textit{Observation of symmetric multicellular WFs.}---A consequence of RTP is the emergence of symmetric multicellular WFs, which cannot adiabatically transform into conventional $\delta$-functions in real space~\cite{nelson2021multicellularity}. This behavior directly results from the large Wannier center variation in RTP. To elucidate this property, we construct the WF by performing the Fourier transform on the experimentally measured Bloch functions of the lower band:
\begin{equation}
{w_{\mathbf{R}}}\left( {\mathbf{r}} \right) = S\int_\text{BZ} {\frac{{d{\mathbf{k}}}}{{{{\left( {2\pi } \right)}^2}}}{e^{i{\mathbf{k}} \cdot \left( {{\mathbf{r}} - {\mathbf{R}}} \right)}}{u_{\mathbf{k}}}\left( {\mathbf{r}} \right){e^{i{\theta _{\mathbf{k}}}}}} ,\label{Wannier}
\end{equation}
where ${w_{\mathbf{R}}}\left( {\mathbf{r}} \right) = {w_{\mathbf{0}}}\left( {{\mathbf{r}} - {\mathbf{R}}} \right)$ is a periodic WF labeled by the Bravais lattice vector $\mathbf{R}$, and $S$ is the unit cell area. Here, $e^{i{\theta _{\mathbf{k}}}}$ represents the gauge freedom that exists in the definition of the Bloch functions and is inherited by the WFs. Due to this gauge freedom, the WFs do not necessarily possess the symmetries of the crystal~\cite{marzari2012maximally, pizzi2020wannier90}. By selecting a special gauge (see \SI), we construct symmetric WFs, which are shown in Fig.~\ref{fig2}c. The results from the tight-binding model, full-wave simulation, and experiment agree well with each other, clearly demonstrating the multicellular feature of delicate topology.

\textit{Observation of gapless edge modes.}---Although the Chern number of a delicate topological insulator is always zero, gapless edge modes can still emerge due to the unique properties of delicate topological bands. Intuitively, the RTP feature in the Wannier band suggests that the Wannier center exhibits significant variations across different momenta. Consequently, the WFs, although localized, are more extended than conventional ones, i.e., multicellular WFs. Therefore, when considering a finite system, the poorly localized WFs may experience separation due to the presence of a boundary, giving rise to boundary modes. Note that this picture differs from the situation in obstructed atomic insulators~\cite{benalcazar2017quantized, benalcazar2017electric}, where edge and corner modes are induced by dividing the centers of well-localized WFs. Moreover, the boundary modes of a delicate topological insulator are usually gapless, while those of an obstructed atomic insulator are commonly gapped.

The gapless edge modes in our delicate topological insulator model can be understood from an sBZ topology perspective~\cite{chen2024chern}. Due to the mirror symmetry $\mathcal{M}_y = \sigma_z$, each half BZ (i.e., $0<\theta<\pi$ and $\pi<\theta<2\pi$) forms a closed manifold and supports a well-defined Chern number. For our acoustic model, we find that, in both theory and experiment, the lower and upper halves of the BZ have opposite Berry curvatures, with their integrations yielding Chern numbers of $-1$ and $+1$, respectively (see Fig.~\ref{fig3}a). Applying Stokes' theorem, we find that:
\begin{equation}
{p_x}\left( {{\theta _1}} \right) - {p_x}\left( {{\theta _2}} \right) = \frac{1}{{2\pi }}\int_0^{2\pi } {d{k_x}} \int_{{\theta _1}}^{{\theta _2}} {d\theta\, } \Omega \left( {\mathbf{k}} \right) \textcolor{cyan}{,}\label{eq06}
\end{equation}
where $\Omega \left( {\mathbf{k}} \right)$ is the Berry curvature. Equation~\eqref{eq06} is consistent with the RTP spectrum, showing opposite quantized windings of bulk polarization in the two halves of the BZ. In each half BZ, the nonzero sBZ Chern number induces a pair of gapless chiral edge modes on opposite edges, propagating in the synthetic dimension $\theta$. The propagation directions of the chiral edge modes are opposite in different halves of the BZ, as enforced by the opposite sBZ Chern numbers in the bulk (see Fig.~\ref{fig3}b). In total, there is one pair of counterpropagating modes on each edge. Unlike other classical realizations of counterpropagating topological modes that do not always span the entire bandgap---such as the analogs of quantum spin Hall~\cite{kane2005quantum, bernevig2006quantum} and valley Hall~\cite{xiao2007valley, mak2014valley} systems---the chiral edge modes here always span the entire bulk bandgap regardless of system parameters due to the protection by the sBZ Chern numbers (see \SI).

To probe the edge modes experimentally, we measure the response of an edge resonator by simultaneously placing the source and the probe in the resonator. This measurement is repeated for all edge resonators across all samples (see \SI). Figures~\ref{fig3}c and \ref{fig3}d depict the measured acoustic pressure amplitude spectra as a function of $\theta$ on the left and right edges, respectively. As observed, each spectrum contains a single peak, corresponding to the edge mode. The peak frequencies match well with the numerical edge dispersions shown in Fig.~\ref{fig3}b, demonstrating one pair of counterpropagating gapless modes per edge. We note that there are no gapless modes on the $x$-directional edge since the Wannier band along $k_x$ does not exhibit the RTP, and the two sBZ Chern numbers cancel each other when projected onto the $k_x$ axis (see Fig.~\ref{fig3}a and \SI).

\textit{Conclusions.}---Going beyond Thouless pumping, we experimentally realize RTP through the observations of bulk polarization and symmetric multicellular WFs. Additionally, we observe gapless edge modes, dictated by the nonzero sBZ Chern numbers. By displaying the features unique to delicate topology, our acoustic system serves as an ideal platform to explore RTP and related properties~\cite{chen2024chern}.

Future studies can leverage synthetic dimensions~\cite{yuan2018synthetic, ozawa2019topological2} to explore more complex delicate topological insulator models and investigate richer forms of RTP by considering other crystalline symmetries. Our system also allows for the incorporation of non-Hermiticity, such as tailored loss via absorptive materials and asymmetric hoppings using active components~\cite{gao2021non, zhang2021acoustic}, opening new avenues in non-Hermitian delicate topology and related RTP. On the practical side, sBZ topology can be further enriched by partitioning the BZ into more than two sBZs with nonzero Chern numbers~\cite{chen2024chern}, leading to multiple gapless edge modes with potential applications in multimode waveguiding and signal multiplexing.


%

\section*{Methods}

\noindent\textbf{Acoustic lattice design.}---For each resonator, $w = 15$ mm, $H = 10$ mm, $h = 4$ mm, and $d_2 = 6$ mm. The lattice constant is $a = 54$ mm. The parameters $d_1$, $l_1$, $l_2$, and $A$ vary in the 12 samples. The volumes of the leftmost and rightmost cavities are tuned to ensure that the on-site potential at the edges is the same as that at the bulk sites. The lengths of the leftmost and rightmost cavities are $\eta_1 l_1$ and $\eta_2 l_2$, respectively. The detailed parameters are listed in \SI.

\noindent\textbf{Numerical simulations.}---All simulations are performed using the acoustic module of COMSOL Multiphysics, which is based on the finite element method. The photosensitive resin used for sample fabrication is set as a hard boundary due to its large impedance mismatch with air. In Fig.~\ref{fig1}d, we first calculate the band structure of a two-cavity acoustic unit cell and compare it with the tight-binding band structure. From there, we extract $t_1$, $t_2$, and $\delta$ from the acoustic lattice. In Fig.~\ref{fig2}, the band structure and Berry phase are calculated based on the two-cavity acoustic unit cell. In Fig.~\ref{fig3}b, a 48-cavity chain is used for each $\theta$, and we calculate the eigenfrequency for each chain. The real sound speed at room temperature is $c_0 = 346$ m/s. The air density is set to be 1.8 $\text{kg/m}^3$.

\noindent\textbf{Sample details.}---The sample is fabricated using 3D printing technology with a fabrication error of 0.1 mm. The width of the wall is 5 mm. We fabricated 12 samples, each containing 24 cavities and several waveguides. For the positive hoppings, i.e., $0 < \theta < \pi$, the waveguides are crossed, while for the negative hoppings, i.e., $\pi < \theta < 2\pi$, the waveguides are parallel to each other. Each cavity includes two holes with radii of 2 mm and 0.8 mm, respectively. The source and probe are inserted through the larger and smaller holes, respectively. These holes are covered by plugs when not in use.

\noindent\textbf{Experimental measurements.}---In the experiment, a broadband sound signal (5 kHz to 7 kHz) is launched from a tube (approximately 3 mm in diameter) that is inserted into the cavity, which acts as a point-like sound source for the wavelength focused here. The pressure at each site is detected by a microphone (Brüel\&Kjær Type 4182) adhered to a steel tube (1.5 mm in diameter and 20 mm in length). The signal is recorded and frequency-resolved using a multi-analyzer system (Brüel\&Kjær 3560 C module). In Fig.~\ref{fig2}, the source is located in the 13th (12th) cavity for samples 4--9 (samples 1--3, 10--12). We measured the amplitude and phase of the sound pressure at all 24 sites for each chain. In Fig.~\ref{fig3}c (Fig.~\ref{fig3}d), the source is positioned in the leftmost (rightmost) cavity, and we only measure the amplitude and phase of the sound pressure at the leftmost (rightmost) site.

\noindent\textbf{Data analysis.}---In Fig.~\ref{fig2}a, we Fourier transform the real-space fields to obtain the momentum space fields. For each $k$, we fit the field to a Lorentzian line shape around the local maxima. The eigenfrequency corresponds to the location of the maximum of the Lorentzian fit, and the eigenfunction is identified as the momentum space wavefunction. In Fig.~\ref{fig2}b, we calculate the Berry phase using Eq.~(\ref{Berry_phase}).

\bigskip
\noindent\textbf{Data availability.}---The simulation files, experimental data are available in the data repository for Nanyang Technological University at this link (URL to be inserted upon publication). Other data supporting this study’s findings are available from the corresponding authors upon reasonable request.

\bigskip
\noindent{\large{\bf{Acknowledgements}}}
Z.C., Y.L., Z.Y., H.T.T., and B.Z. are supported by the Singapore National Research Foundation Competitive Research Program (Grant No. NRF-CRP23-2019-0007), and the Singapore Ministry of Education Academic Research Fund Tier 2 (Grant No. MOE-T2EP50123-0007) and Tier 1 (Grant Nos. RG139/22 and RG81/23). W.X. and H.X. are supported by the National Natural Science Foundation of China under grant No. 62401491 and the Chinese University of Hong Kong under grants No. 4937205 and 4937206. S.Y., and Y.Z. are supported by the GRF of Hong Kong (Grant No. 17301224).

\bigskip
\noindent{\large{\bf{Author contributions}}}
H.X. conceived the idea. Z.C., S.Y., Y.L., W.X., Y.Z., and H.X. did the theoretical analysis. Z.C. performed the simulations and designed the sample. Z.C., Z.Y., H.H.T. conducted the experiments. Z.C., Y.Z., H.X., and B.Z. wrote the manuscript with input from all authors. Y.Z., H.X., and B.Z. supervised the project. 

\bigskip
\noindent{\large{\bf{Competing interests}}}
The authors declare no competing interests.

\end{document}